\documentclass[10pt,conference]{IEEEtran}
\IEEEoverridecommandlockouts

\usepackage{hyperref}
\usepackage{cite}      
\usepackage{graphicx}  
\usepackage{subfigure} 
\usepackage{url}
\usepackage{stfloats}  
\usepackage{amsmath}   
\usepackage{latexsym}
\usepackage{algorithm,algorithmic}
\usepackage{amssymb}
\usepackage{amsmath}  
\usepackage{amsfonts} 
\usepackage{verbatim}
\usepackage{multirow}
\usepackage{float}

\usepackage[left=1.62cm,right=1.62cm,top=2.1cm,bottom=2.66cm]{geometry}

\begin{document}

\title{BALANCE: Bitrate-Adaptive Limit-Aware Netcast Content Enhancement Utilizing QUBO and Quantum Annealing 
\\\thanks{\textsuperscript{*}Animesh Rajpurohit and Michael Kelley are co-first authors and contributed equally to this work.}}

\vspace{-5mm}
\author{
  \IEEEauthorblockN{
    Animesh Rajpurohit$^\dagger$\textsuperscript{*}, 
    Michael Kelley$^\dagger$$^\dagger$\textsuperscript{*}, 
    Wei Wang$^\dagger$, 
    and Krishna Murthy Kattiyan Ramamoorthy$^{\ddagger}$}
  
  \IEEEauthorblockA{
    \textit{$^\dagger$Department of Computer Science, San Diego State University,} \\
    \textit{$^\dagger$$^\dagger$Department of Astronomy, San Diego State University,} \\
    \textit{$^{\ddagger}$Department of Computer Science and Engineering, Santa Clara University,}\\
    arajpurohit6174@sdsu.edu, mkelley3459@sdsu.edu, wwang@sdsu.edu, kkattiyanramamoorthy@scu.edu}

    \vspace{-10mm}
}\maketitle

\textbf{\textit{Abstract: }}\textbf{In an era of increasing data cap constraints, optimizing video streaming quality while adhering to user-defined data caps remains a significant challenge. This paper introduces Bitrate-Adaptive Limit-Aware Netcast Content Enhancement (BALANCE), a novel Quantum framework aimed at addressing this issue. BALANCE intelligently pre-selects video segments based on visual complexity and anticipated data consumption, utilizing the Video Multimethod Assessment Fusion (VMAF) metric to enhance Quality of Experience (QoE). We compare our method against traditional bitrate ladders used in Adaptive Bitrate (ABR) streaming, demonstrating a notable improvement in QoE under equivalent data constraints. We compare the Slack variable approach with the Dynamic Penalization Approach (DPA) by framing the bitrate allocation problem through Quadratic Unconstrained Binary Optimization (QUBO) to effectively enforce data limits. Our results indicate that the DPA consistently outperforms the Slack Variable Method, delivering more valid and optimal solutions as data limits increase. This new quantum approach significantly enhances streaming satisfaction for users with limited data plans.}


\section{Introduction}
Since the launch of YouTube in 2005, video streaming has become one of the most popular online activities, with users spending approximately 29 hours per month on the platform globally \cite{youtube2024}. As demand for high-quality video content continues to surge, delivering reliable and efficient streaming experiences is crucial, particularly for consumers with limited data plans. Traditional video streaming methods rely on Adaptive Bitrate (ABR) technology \cite{ABR} to adjust to fluctuations in bandwidth, often resulting in suboptimal data consumption. This technique is often inefficient for users, as they are forced to manually lower video quality to conserve data, sacrificing their Quality of Experience (QoE) \cite{QoE}.

This paper aims to address this issue by proposing a novel solution, \textbf{Bitrate-Adaptive Limit-Aware Netcast Content Enhancement (BALANCE)}, which optimizes both video quality and data usage. BALANCE solves a Knapsack optimization problem\cite{Knapsack} to pre-select the best combination of video segments, maximizing the Video Multimethod Assessment Fusion (VMAF)\cite{VMAF_Metric} score while adhering to user-defined data limits. While traditional ABR methods focus on bandwidth adaptation, BALANCE introduces a preemptive approach, considering data capacity and content complexity to enhance the streaming QoE without exceeding data caps. Our proposed BALANCE framework extends the typical ABR pipeline by introducing a quantum-based pre-selection layer that chooses segment bitrates under data constraints. By leveraging pre-encoded video segments, BALANCE seamlessly integrates into existing ABR systems and computes QoE for these segments before playback, enabling a novel approach that respects data limits while optimizing streaming quality, going beyond traditional bandwidth-adaptive decisions.

\enlargethispage{0.015in}

In the current Noisy Intermediate-Scale Quantum (NISQ) era, quantum computers, although not fault-tolerant, have shown promise in efficiently solving combinatorial optimization problems\cite{NISQ}. The NISQ devices exploit quantum phenomena like superposition and entanglement to explore solution spaces more effectively. In particular, combinatorial optimization problems like Knapsack can be framed as Quadratic Unconstrained Binary Optimization (QUBO) problems, which can be efficiently addressed using quantum annealing methods or hybrid quantum-classical algorithms \cite{lucas2014ising}.

Quantum annealers have demonstrated the potential to outperform classical systems in solving complex optimization problems in various domains, including quantum machine learning \cite{Quantum_Use_Machine}, financial portfolio optimization \cite{Quantum_Use_financial} and logistics \cite{Quantum_Use_Logistics_1}. By leveraging quantum computing's unique capabilities, particularly in the NISQ era, we aim to optimize video streaming for users with limited data plans, by providing enhanced video quality and a superior QoE within their data limits.


In this paper, we propose integrating quantum approaches, particularly QUBO formulations, into the ABR framework to address the video bitrate allocation. We demonstrate the potential of quantum-based optimization using quantum annealing to handle large-scale, real-world data challenges. Building on existing research, this work advances the application of quantum computing to practical problems, aiming to achieve a ‘quantum advantage’ in industries like multimedia streaming.

\enlargethispage{0.015in}

\vspace{-1mm}
\section{Background and Related Work}
\vspace{-2mm}

\subsection{Background}
\subsubsection{Adaptive Bitrate (ABR)}
ABR streaming is a widely adopted technology that enables dynamic adjustment of video quality based on users network conditions. The ABR algorithm breaks a video into various segments and encodes each segment with different bitrates \cite{Deep_Reinforced_Bitrate_Ladders_for_ABR}. Each encoded segment requires varying bandwidth based on its quality. The algorithm selects a quality (e.g., 360p/480p/720p) that matches the available bandwidth but does not account for total data usage, focusing solely on current network conditions.

\subsubsection{Quality of Experience (QoE)}

QoE evaluates user satisfaction during service consumption \cite{QoE2} and is measured using VMAF, an objective metric representing human-perceived quality \cite{VMAF_Metric}. VMAF, preferred over PSNR and SSIM for its superior representation of perceived quality \cite{VMAF_BL, VMAF2}, guides encoding and bitrate ladder design. However, ABR, while aimed at improving QoE, struggles to maximize it efficiently with available data.


\subsubsection{Quadratic Unconstrained Binary Optimization (QUBO)}


QUBO is a mathematical framework for modeling combinatorial optimization problems as a quadratic objective function \cite{Formulating_QUBO}, with binary decision variables and quadratic interactions. Its goal is to find variable configurations that minimize or maximize the objective function. The general form is:

\vspace{-2mm}
\begin{equation}
f(x) = \sum_{i} q_{ii} x_i + \sum_{i<j} q_{ij} x_i x_j
\vspace{-2mm}
\end{equation}

\noindent where \(x_i\) represents binary variables (0 or 1), \(q_{ii}\) denotes linear coefficients representing the contribution of individual variables, and \(q_{ij}\) indicates quadratic coefficients representing the interactions between pairs of variables. This quadratic form can be transformed into the equivalent Ising model, where the binary variables \(x_i\) are mapped to spin variables that take values of either \( -1 \) or \( +1 \). The Ising model is widely used in quantum computing, as it simplifies the quantum states\cite{lucas2014ising} needed for problem formulation.

\subsubsection{Quantum Annealing}


Quantum annealing is a quantum computational technique for solving optimization problems like those in the QUBO framework \cite{Perspectives_Of_QA}. It uses quantum phenomena such as superposition and tunneling to explore solution spaces and find the global minimum efficiently \cite{Quantum_Annealing_Textbook}. QUBO problems are mapped to quantum systems, where qubits represent binary variables, and the system evolves by minimizing energy to reach the optimal solution. Unlike classical algorithms, quantum annealing avoids local minima through tunneling, making it effective for optimizing video segment parameters to maximize QoE within users constraints.

\vspace{3mm}
\subsection{Related Work}

Huang et al. \cite{Deep_Reinforced_Bitrate_Ladders_for_ABR} proposed DeepLadder, a deep reinforcement learning-based system that optimizes bitrate ladders for adaptive video streaming by considering video content features, network capacity, and storage costs, achieving significant improvements in video quality, bandwidth utilization, and storage efficiency over traditional methods. Chen et al. \cite{Optimized_Transcoding_for_Large_Scale_Adaptive_Streaming_Using_Playback_Statistics} introduced a method that uses empirical data on bandwidth and viewport size distributions to optimize video encoding bitrates, reducing average bitrate by 9.7\% without compromising quality. Juluri et al. \cite{SARA:_Segment_Aware_Rate_Adaptation_Algorithm_for_Dynamic_Adaptive_Streaming_over_HTTP} developed the Segment-Aware Rate Adaptation (SARA) algorithm for DASH, enhancing video playback quality by incorporating segment size information for accurate download time predictions and improved QoE. BALANCE builds on these works by tackling the under explored challenge of optimizing video quality under user-defined data caps. While DeepLadder optimizes bandwidth usage and SARA improves bitrate adaptation using segment size information, neither explicitly manages total data consumption for users with capped plans. BALANCE addresses this gap by formulating the bitrate allocation problem as a Quadratic Unconstrained Binary Optimization (QUBO) and leveraging quantum annealing to preemptively manage data allocation while enhancing QoE. This novel approach integrates seamlessly with existing ABR systems, offering a data-aware optimization designed for users with limited data plans.


Jiang and Chu \cite{Classify_QUBO_NP} explore solving NP-hard problems with Quantum Annealing Algorithms (QAAs) based on QUBO, introducing a classification system with four classes defined by variable scaling and the presence of constraint and optimization terms. They benchmark QAAs on a D-Wave quantum annealer, comparing them to classical algorithms and identifying cases where QAAs excel in solving complex problems. Montañez-Barrera et al. \cite{Main_Paper} propose reducing slack variables through unbalanced penalization, simplifying problem representation and enhancing QAA performance for combinatorial optimization tasks like the Traveling Salesman Problem (TSP). Ikeda et al. \cite{Nurse_Scheduling_QA} demonstrate quantum annealing’s practical application using the D-Wave 2000Q to solve the Nurse Scheduling Problem (NSP), highlighting its ability to efficiently handle hard constraints and improve workforce scheduling in real-world scenarios.

\begin{figure*}[t]
\vspace{4mm}
\begin{center}
  \includegraphics[width=.9\textwidth, height=0.2\textheight]{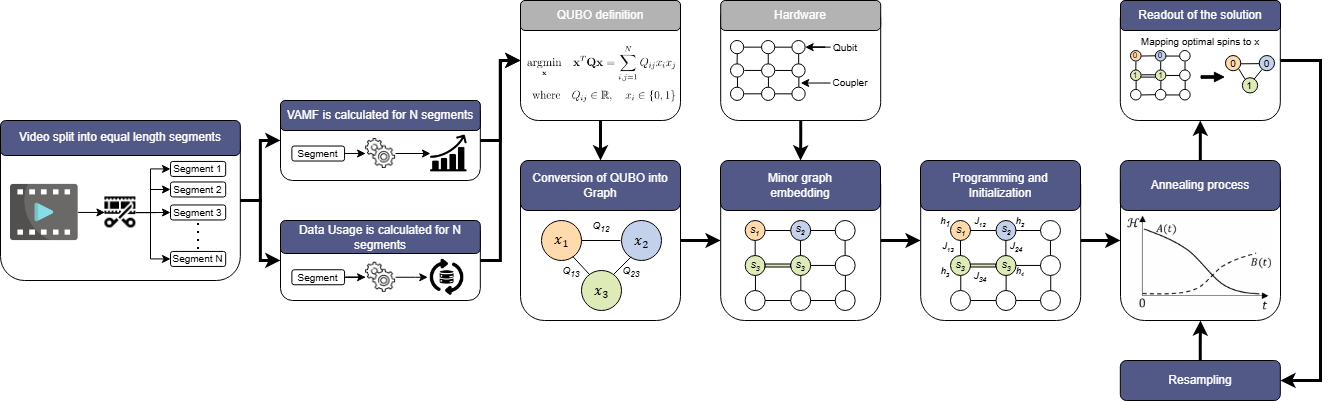}
\end{center}
\vspace{-4mm}
\caption{The flowchart illustrates the overview of the BALANCE process. This figure is adapted from \cite{Classify_QUBO_NP}.}
\label{Flow_total}
\vspace{-6mm}
\end{figure*}

\section{Problem Formulation}

\subsection{Bitrate-Adaptive Limit-Aware Netcast Content Enhancement (BALANCE)}




In this study, two key analyses were conducted on each video. First, the videos were broken up into equal-length segments, ranging from two to six segments. All segments were then encoded with varying bitrates and their quality was assessed using VMAF. Second, the total data usage for each segment during encoding was recorded, enabling an evaluation of trade-offs between video quality and data consumption. 

To assess the quality and data usage across different encoding settings, the clips were encoded at four different bitrates: 8000kbps, 5000kbps, 2500kbps, 1000kbps. The encoding parameters were based on the recommended video bitrates by YouTube for standard dynamic range (SDR) uploads with a standard frame rate \cite{youtube_encoding_settings}, ensuring consistency across videos. For quality assessment, VMAF scores were calculated by comparing each encoded segment to its corresponding reference segment. These results provide valuable insights into how the varying encoding parameters impact visual quality and data consumption, contributing to our QUBO model.


Fig. \ref{Flow_total} provides a detailed overview of the steps involved in optimizing video quality and data usage through quantum annealing. The QUBO problem is formulated and then mapped onto the quantum processor through minor embedding. After initialization, the annealing process is executed to find the optimal solution, which is read out and post-processed to derive the final optimized solution.

\subsection{QUBO Formulation for Video Streaming Optimization}

We formulate the video streaming optimization problem using QUBO to maximize video quality while adhering to data usage constraints. The optimization involves selecting the optimal quality level for each video segment, where binary decision variables represent the chosen quality for each segment.

\noindent \textit{Problem Definition:} Let \( x_{i,j} \in \{0, 1\} \) be a binary variable, where \( x_{i,j} = 1 \) if segment \( i \) is encoded at quality level \( j \), and \( x_{i,j} = 0 \) otherwise. Here, \( v_{i,j} \) is the VMAF score, and \( d_{i,j} \) is the data usage for segment \( i \) at quality level \( j \). With \( N \) segments and \( M \) quality levels, the objective is to maximize the total VMAF score:

\vspace{-3mm}
\begin{equation} 
\text{Maximize} \quad \sum_{i=1}^{N} \sum_{j=1}^{M} v_{i,j} x_{i,j} \label{eq:obj} 
\end{equation}
\vspace{-2mm}

\noindent subject to the selection and data usage constraints. 

\noindent \textit{Selection Constraint:} Each video segment must have exactly one quality level selected. This is represented by the equality constraint:

\vspace{-2mm}
\begin{equation} \sum_{j=1}^{M} x_{i,j} = 1 \quad \forall i \in {1, \dots, N} 
\end{equation}
\vspace{-2mm}

QUBO does not naturally accept equality constraints \cite{Formulating_QUBO}, and so we encode this constraint using a penalty term. The penalty ensures that any solution where more than one or no quality levels being selected for a segment is discouraged, driving the optimization towards valid solutions:

\vspace{-2mm}
\begin{equation} 
-\lambda_0 \sum_{i=1}^{N} \left( \sum_{j=1}^{M} x_{i,j} - 1 \right)^2 \label{eq:equality} 
\end{equation}
\vspace{-2mm}

\noindent where $\lambda_0$ is the penalization coefficient. 

\noindent \textit{Data Usage Constraint:} The total data usage across all video segments must remain within the predefined limit $D_{\text{max}}$:
\begin{equation} \sum_{i=1}^{N} \sum_{j=1}^{M} d_{i,j} x_{i,j} \leq D_{\text{max}} \end{equation}
\vspace{-2mm}

QUBO also does not directly accept inequality constraints \cite{Formulating_QUBO}, we introduce two methods to handle this constraint: the Slack Variable Method and the Dynamic Penalization Approach.

\paragraph{\textbf{Slack Variable Method}}
To convert the inequality into an equality, we introduce a slack variable $S$ that satisfies:
\begin{equation} \sum_{i=1}^{N} \sum_{j=1}^{M} d_{i,j} x_{i,j} + S = D_{\text{max}} \end{equation}

\noindent $S\geq 0$ and can be expressed as a sum of binary slack variables:
\begin{equation} S = \sum_{k=1}^{K} 2^{k-1} s_k \
\end{equation}
\vspace{-2mm}

\noindent $s_k \in \{0,1\}$ are binary slack variables, and $K$ is the number of slack variables needed, where 
\begin{equation} K = \lceil \log_2(D_{\text{max}}) \rceil \end{equation}

\noindent This introduces an additional penalization term for the data usage constraint:
\begin{equation} -\lambda_1 \left( D_{\text{max}} - \sum_{i=1}^{N} \sum_{j=1}^{M} d_{i,j} x_{i,j} - \sum_{k=1}^{K} 2^{k-1} s_k \right)^2 
\label{eq:slack}
\end{equation}

\noindent where $\lambda_1$ is the penalization coefficient enforcing this constraint. The choice of $\lambda_1$ reflects the trade-off between adhering to data usage constraints and maximizing video quality, ensuring neither aspect dominates disproportionately.

\paragraph{\textbf{Dynamic Penalization Approach (DPA)}}

The DPA is a heuristic method, inspired by the Unbalanced Penalization approach\cite{Main_Paper}, designed to adaptively enforce the data usage constraint. Unlike the slack variable method, which introduces additional variables, or unbalanced penalization, which requires careful selection of $\lambda$-values for different limits, DPA offers flexibility by dynamically adjusting the penalization without requiring re-calibration of penalty coefficients.

DPA applies a more gradual penalization approach, where the penalty begins before reaching the data limit $D_{\text{max}}$. The penalization starts at a threshold
\begin{equation} D_{\text{threshold}} = \frac{D_{\text{max}}}{\mu_3} \end{equation}

\noindent where $\mu_3 > 1$ controls how early the penalty begins.

The penalization term is defined as:
\begin{align}
+&\mu_1 \left( \frac{\sum_{i=1}^{N} \sum_{j=1}^{M} d_{i,j} x_{i,j} - D_{\text{threshold}}}{D_{\text{max}} - D_{\text{threshold}}} \right) \nonumber \\
-& \mu_2 \left( \frac{\sum_{i=1}^{N} \sum_{j=1}^{M} d_{i,j} x_{i,j} - D_{\text{threshold}}}{D_{\text{max}} - D_{\text{threshold}}} \right)^2
\label{eq:dpa}
\end{align}

\noindent where $\mu_1$ controls the linear penalization growth and $\mu_2$ controls the quadratic penalization growth.

DPA ensures smoother constraint enforcement by gradually increasing the penalty as the solution approaches the data usage limit, offering more flexibility without introducing extra slack variables. The final QUBO formulation referring from Eq.\eqref{eq:obj} and Eq.\eqref{eq:equality} can be summarized as follows:

\begin{equation} \text{Max} \left[ \sum_{i=1}^{N} \sum_{j=1}^{M} v_{i,j} x_{i,j} - \lambda_0 \sum_{i=1}^{N} \left( \sum_{j=1}^{M} x_{i,j} - 1 \right)^2 + \text{InEqPen} \right] 
\label{eq:qubo}
\end{equation}

The InEqPen (Inequality Penalty) in Eq.\eqref{eq:qubo}  can be described using either Eq.\eqref{eq:slack} which follows the Slack Variable Method, or Eq.\eqref{eq:dpa}, which applies the DPA.

\section{Simulation Study And Results Discussion}

\subsection{QUBO Model and Energy Landscape}
\begin{figure}[H]
\centering
\includegraphics[width=0.45\textwidth]{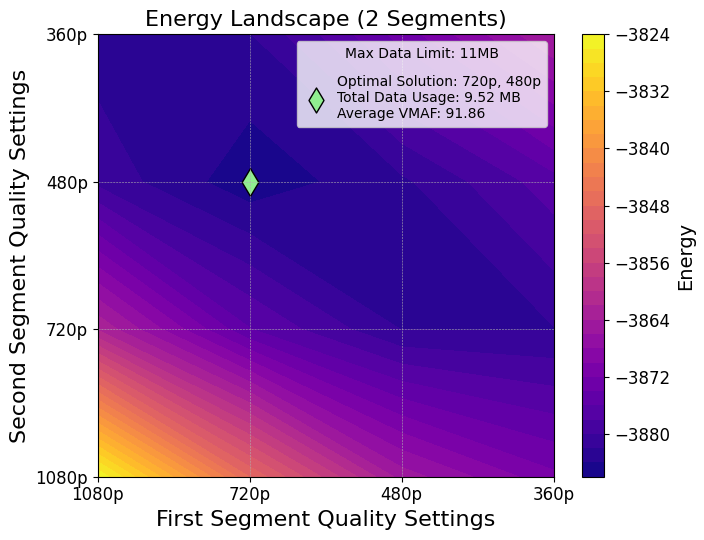}
\vspace{-3mm}
\caption{The 2D energy landscape shows energy variation with quality settings for two video segments. Contour lines highlight equal energy regions, indicating optimal solutions that maximize VMAF within data constraints.}
\label{fig:fig2}
\vspace{-3mm}
\end{figure}

To solve the bitrate allocation problem efficiently, we utilized QUBO. The optimization was modeled using two methods for handling data constraints: the Slack Variable Method and the DPA. The problem was modeled using IBM's Docplex and converted to QUBO using \textit{OpenQAOA}, an SDK designed for quantum optimization\cite{openqaoasdkqaoa}. The resulting QUBO models were executed on the \textit{D-Wave Advantage} quantum annealer, which leverages quantum tunneling and superposition to efficiently explore the solution space\cite{dwave_documentation}. To visualize the QUBO problem, a 2D contour plot of the energy landscape was generated for a two-segment video with different quality settings, i.e. 360p/480p/720p/1080p (Fig. \ref{fig:fig2}).

The energy landscape shown in Fig. \ref{fig:fig2} illustrates the cost function for combinations of quality settings, where lower energy values represent more optimal solutions. 
The DPA method consistently guides the system toward these low-energy regions, helping avoid local minima. Table \ref{table:QUBO_Energy_Landscape} presents the VMAF scores and data usage for different resolution settings across the two video segments. These values form the basis for calculating the cost function in the QUBO model. 

\begin{table}[h!]
\centering
\caption{VMAF and Data Usage at Different Resolutions}
\label{table:segment_data}
\begin{tabular}{|c|c|c|c|c|}
\hline
\textbf{Segment} & \textbf{Resolution} & \textbf{VMAF} & \textbf{Data Usage (MB)} \\ \hline
\multirow{4}{*}{First Segment} & 1080p & 92.90 & 8.17 \\ \cline{2-4} 
                               & 720p  & 90.58 & 5.46 \\ \cline{2-4} 
                               & 480p  & 87.13 & 2.68 \\ \cline{2-4} 
                               & 360p  & 84.65 & 0.96 \\ \hline
\multirow{4}{*}{Second Segment} & 1080p & 95.69 & 12.09 \\ \cline{2-4} 
                               & 720p  & 94.96 & 7.76 \\ \cline{2-4} 
                               & 480p  & 93.14 & 4.06 \\ \cline{2-4} 
                               & 360p  & 89.03 & 1.63 \\ \hline
\end{tabular}
\label{table:QUBO_Energy_Landscape}
\vspace{-5mm}
\end{table}

\subsection{Video Quality and Data Usage Analysis}

To evaluate the effectiveness of the BALANCE framework, we utilized three distinct datasets, each selected to test specific aspects of the methodology and allow for a comprehensive validation of BALANCE across a range of scenarios:

\begin{itemize}
    \item \textbf{YouTube User Generated Content (UGC) Dataset}~\cite{UGCDataset}: This dataset consists of videos across 15 categories of user-generated content, representing diverse resolutions, encoding complexities, and scene dynamics. It is a representative benchmark for evaluating typical consumer-level streaming scenarios.
    \item \textbf{Netflix Public Dataset}~\cite{NetflixDataset}: This dataset contains professionally produced videos with high motion complexity, varied lighting, and a wide range of resolutions. It provides a robust test case for the application of BALANCE on complex, high-quality streaming content.
    \item \textbf{AVA Actions Dataset}~\cite{AVADataset}: This dataset contains 15-minute videos characterized by diverse motion patterns and high temporal complexity. It is used to evaluate BALANCE's scalability and performance in optimizing quality and data usage for extended, motion-intensive scenarios.
\end{itemize}

\begin{figure}[H]
\begin{center}
  \includegraphics[width=.45\textwidth]{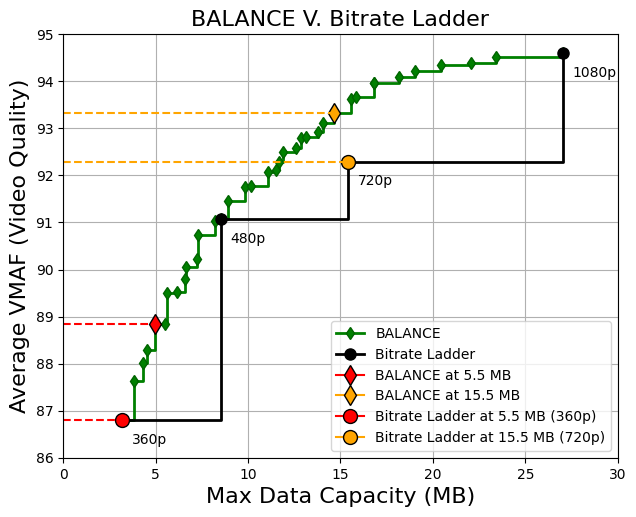}
\end{center}
\caption{Comparison of BALANCE and traditional bitrate ladders for a 20-second video with 4-second segments. BALANCE achieves higher VMAF scores and better QoE while adhering to data constraints, outperforming static ladders at equivalent data caps.}

\label{5_segment}
\vspace{-3mm}
\end{figure}

In the simulation study, the BALANCE approach demonstrates two key benefits over traditional bitrate ladder methods. Fig. \ref{5_segment}  illustrates a scenario where a user has a data limit that falls between the requirements for 360p and 480p. Instead of forcing the user to stick to a single lower quality setting, like 360p, for the entire video, BALANCE effectively selects a combination of qualities for the segments achieving a broader range of higher average VMAF scores while staying within the user's data limit.

Additionally, BALANCE can deliver higher visual quality while using less data. As shown in Fig. \ref{5_segment}, during transitions such as 480p to 720p, BALANCE achieves a higher average VMAF score with lower data consumption by selecting segments with minimal data increase to be displayed at a higher quality, while saving data on less complex segments without significantly reducing VMAF. This approach allows for a more efficient use of data, delivering a higher overall QoE. Furthermore, BALANCE allows users to set specific data caps, and the system allocates data across viewing hours to maximize QoE without exceeding data limits. By optimizing the quality across segments, BALANCE maximizes visual quality while adhering to data constraints, offering a more refined and flexible streaming experience, as evidenced by our results.

\subsection{Key Findings on DPA vs. Slack Variables}

The performance of the QUBO models was evaluated based on the probability of finding valid and optimal solutions. Valid solutions satisfy both the selection and data usage constraints, while optimal solutions further maximize the VMAF score.

\begin{figure}[!ht]
\begin{center}
  \includegraphics[width=.45\textwidth]{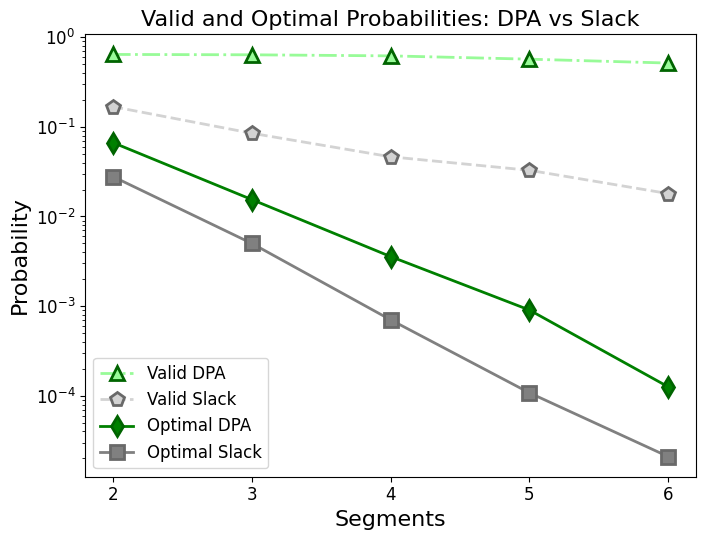}
\end{center}
\caption{The probability of valid and optimal solutions for DPA vs. Slack Variable Method across segments, executed on D-Wave Advantage QPU (1000 shots/run, averaged over 10 trials).}
\label{fig:QPU_Counts}
\vspace{-3mm}
\end{figure}

Fig. \ref{fig:QPU_Counts} summarizes the performance results, showing the probability of valid and optimal solutions across different data capacities for both the DPA and the Slack Variable Method. The DPA approach consistently shows higher probabilities of achieving valid and optimal solutions compared to the Slack Variable Method.
The results indicate that the DPA consistently outperforms the Slack Variable Method, particularly as the data capacity increases. This is demonstrated by the higher probabilities of both valid and optimal solutions in the DPA approach. As data capacity increases, the probability of obtaining valid and optimal solutions improves more significantly with DPA compared to the Slack Variable Method, which struggles to find optimal solutions, especially at higher capacities.

\subsection{Optimizing DPA Parameters}

The constant parameters used in the Dynamic Penalization Approach (DPA) were optimized through an iterative heuristic approach to maximize the proportion of valid and optimal solutions obtained from the quantum annealer. Specifically, a range of values for $\mu_1$, $\mu_2$, and $\mu_3$ were systematically tested by solving the QUBO on the quantum annealer and recording the outcomes. During each iteration, the proportion of solutions satisfying all constraints (valid solutions) and those achieving the highest VMAF scores within the specified data cap (optimal solutions) were evaluated. The parameter combination that consistently yielded the highest percentage of valid and optimal solutions across multiple runs was selected as the final configuration.

The resulting optimal values are summarized in Table \ref{table:Constants}. These constants represent the best-performing configuration, effectively balancing linear and quadratic penalization. Their impact is demonstrated in Figure \ref{fig:QPU_Counts}, which highlights the effectiveness of the DPA method in comparison to the Slack Variable approach.

\begin{table}[!ht]
\vspace{-3mm}
\centering
\caption{Optimized Constant Values for DPA}
\label{tab:constants}
\begin{tabular}{|c|c|}
\hline
\textbf{Constant} & \textbf{Value} \\ \hline
$\mu_1$ (Linear penalization growth)  & 5.6  \\ \hline
$\mu_2$ (Quadratic penalization growth) & 8.9  \\ \hline
$\mu_3$ (Penalty start threshold)  & 1.69 \\ \hline
\end{tabular}
\label{table:Constants}
\vspace{-2mm}
\end{table}


\section{Conclusions}

In this paper, we introduced BALANCE, a quantum-based framework that optimizes video streaming Quality of Experience (QoE) under strict data constraints by pre-selecting video segments based on visual complexity and data usage. Using the Quadratic Unconstrained Binary Optimization (QUBO) model and quantum annealing, BALANCE addresses the NP-hard problem of bitrate allocation, outperforming traditional Adaptive Bitrate (ABR) methods, which focus solely on bandwidth adaptation. Among penalty-based methods, the Dynamic Penalization Approach (DPA) achieved superior results over Slack Variables, yielding more valid and optimal solutions under various data limits. Scalability and real-time applicability remain key challenges due to the increasing computational demands of QUBO models on Noisy Intermediate-Scale Quantum (NISQ) hardware. Practical deployments could benefit from strategies like pre-computing segment priorities or integrating hybrid classical-quantum solvers, where quantum annealing excels in complex scenarios by escaping local minima. While current quantum hardware faces limitations, advancements in qubit scalability and connectivity enhance BALANCE’s potential for large-scale video optimization, laying the groundwork for future research into hybrid approaches, advanced penalty mechanisms, and scalable methods to enable next-generation streaming solutions.

\vspace{-2mm}

\section{Acknowledgements}
The authors thank Dr. J. A. Montañez-Barrera, Postdoctoral Researcher at Forschungszentrum Jülich, for his valuable suggestions and insights.

\vspace{-2mm}


\bibliographystyle{IEEEtran}  
\bibliography{references}  

\begin{thebibliography}{10}
\providecommand{\url}[1]{#1}
\csname url@samestyle\endcsname
\providecommand{\newblock}{\relax}
\providecommand{\bibinfo}[2]{#2}
\providecommand{\BIBentrySTDinterwordspacing}{\spaceskip=0pt\relax}
\providecommand{\BIBentryALTinterwordstretchfactor}{4}
\providecommand{\BIBentryALTinterwordspacing}{\spaceskip=\fontdimen2\font plus
\BIBentryALTinterwordstretchfactor\fontdimen3\font minus \fontdimen4\font\relax}
\providecommand{\BIBforeignlanguage}[2]{{%
\expandafter\ifx\csname l@#1\endcsname\relax
\typeout{** WARNING: IEEEtran.bst: No hyphenation pattern has been}%
\typeout{** loaded for the language `#1'. Using the pattern for}%
\typeout{** the default language instead.}%
\else
\language=\csname l@#1\endcsname
\fi
#2}}
\providecommand{\BIBdecl}{\relax}
\BIBdecl

\bibitem{youtube2024}
{DataReportal, We Are Social, and Meltwater}, ``{Monthly time spent on the YouTube mobile app per user in selected markets worldwide as of July 2024 (in hours) [Graph]},'' \url{https://www.statista.com/statistics/1287283/time-spent-youtube-app-selected-countries/}, 2024.

\bibitem{ABR}
Y.~Sani, A.~Mauthe, and C.~Edwards, ``Adaptive bitrate selection: A survey,'' \emph{IEEE Communications Surveys \& Tutorials}, vol.~19, no.~4, pp. 2985--3014, 2017.

\bibitem{QoE}
A.~A. Barakabitze, N.~Barman, A.~Ahmad, S.~Zadtootaghaj, L.~Sun, M.~G. Martini, and L.~Atzori, ``Qoe management of multimedia streaming services in future networks: A tutorial and survey,'' \emph{IEEE Communications Surveys \& Tutorials}, vol.~22, no.~1, pp. 526--565, 2020.

\bibitem{Knapsack}
M.~Assi and R.~A. Haraty, ``A survey of the knapsack problem,'' in \emph{2018 International Arab Conference on Information Technology (ACIT)}, 2018, pp. 1--6.

\bibitem{VMAF_Metric}
J.~Zhu, H.~Amirpour, R.~Schatz, C.~Timmerer, and P.~L. Callet, ``Enhancing satisfied user ratio (sur) prediction for vmaf proxy through video quality metrics,'' in \emph{2023 IEEE International Conference on Visual Communications and Image Processing (VCIP)}, 2023, pp. 1--5.

\bibitem{NISQ}
J.~Preskill, ``Quantum {C}omputing in the {NISQ} era and beyond,'' \emph{{Quantum}}, vol.~2, p.~79, Aug. 2018.

\bibitem{lucas2014ising}
A.~Lucas, ``Ising formulations of many np problems,'' \emph{Frontiers in Physics}, vol.~2, p.~5, 02 2014.

\bibitem{Quantum_Use_Machine}
E.~H. Houssein, Z.~Abohashima, M.~Elhoseny, and W.~M. Mohamed, ``Machine learning in the quantum realm: The state-of-the-art, challenges, and future vision,'' \emph{Expert Systems with Applications}, vol. 194, p. 116512, 2022.

\bibitem{Quantum_Use_financial}
G.~Rosenberg, P.~Haghnegahdar, P.~Goddard, P.~Carr, K.~Wu, and M.~Lopez~de Prado, ``Solving the optimal trading trajectory problem using a quantum annealer,'' 08 2015.

\bibitem{Quantum_Use_Logistics_1}
H.~Jiang, M.~Shen, and J.~Liu, ``Quantum computing methods for supply chain management,'' 12 2022, pp. 400--405.

\bibitem{Deep_Reinforced_Bitrate_Ladders_for_ABR}
T.~Huang, R.-X. Zhang, and L.~Sun, ``Deep reinforced bitrate ladders for adaptive video streaming,'' in \emph{Proceedings of the 31st ACM Workshop on Network and Operating Systems Support for Digital Audio and Video}, ser. NOSSDAV '21.\hskip 1em plus 0.5em minus 0.4em\relax New York, NY, USA: Association for Computing Machinery, 2021, p. 66–73.

\bibitem{QoE2}
K.~M.~K. Ramamoorthy and W.~Wang, ``Qoe-sensitive economic pricing model for wireless multimedia communications using stackelberg game,'' in \emph{2019 IEEE Global Communications Conference (GLOBECOM)}, 2019, pp. 1--6.

\bibitem{VMAF_BL}
A.~Katsenou, F.~Zhang, K.~Swanson, M.~Afonso, J.~Sole, and D.~Bull, ``Vmaf-based bitrate ladder estimation for adaptive streaming,'' 06 2021, pp. 1--5.

\bibitem{VMAF2}
E.~Ballesteros, K.~M.~K. Ramamoorthy, and W.~Wang, ``Exploring av1 encoder potentials for priority-driven wireless multimedia services,'' in \emph{2022 Intermountain Engineering, Technology and Computing (IETC)}, 2022, pp. 1--6.

\bibitem{Formulating_QUBO}
F.~Glover, G.~Kochenberger, R.~Hennig, and Y.~Du, ``Quantum bridge analytics i: a tutorial on formulating and using qubo models,'' \emph{Annals of Operations Research}, vol. 314, 07 2022.

\bibitem{Perspectives_Of_QA}
P.~Hauke, H.~Katzgraber, W.~Lechner, H.~Nishimori, and W.~Oliver, ``Perspectives of quantum annealing: Methods and implementations,'' \emph{Reports on Progress in Physics}, vol.~83, 05 2020.

\bibitem{Quantum_Annealing_Textbook}
C.~Mcgeoch, ``Adiabatic quantum computation and quantum annealing: Theory and practice,'' \emph{Synthesis Lectures on Quantum Computing}, vol.~5, pp. 1--93, 07 2014.

\bibitem{Optimized_Transcoding_for_Large_Scale_Adaptive_Streaming_Using_Playback_Statistics}
C.~Chen, Y.-C. Lin, S.~Benting, and A.~Kokaram, ``Optimized transcoding for large scale adaptive streaming using playback statistics,'' in \emph{2018 25th IEEE International Conference on Image Processing (ICIP)}, 2018, pp. 3269--3273.

\bibitem{SARA:_Segment_Aware_Rate_Adaptation_Algorithm_for_Dynamic_Adaptive_Streaming_over_HTTP}
P.~Juluri, V.~Tamarapalli, and D.~Medhi, ``Sara: Segment aware rate adaptation algorithm for dynamic adaptive streaming over http,'' in \emph{2015 IEEE International Conference on Communication Workshop (ICCW)}, 2015, pp. 1765--1770.

\bibitem{Classify_QUBO_NP}
J.-R. Jiang and C.-W. Chu, ``Classifying and benchmarking quantum annealing algorithms based on quadratic unconstrained binary optimization for solving np-hard problems,'' \emph{IEEE Access}, vol.~11, pp. 104\,165--104\,178, 2023.

\bibitem{Main_Paper}
J.~A. Montañez-Barrera, P.~van~den Heuvel, D.~Willsch, and K.~Michielsen, ``Improving performance in combinatorial optimization problems with inequality constraints: An evaluation of the unbalanced penalization method on d-wave advantage,'' in \emph{2023 IEEE International Conference on Quantum Computing and Engineering (QCE)}, vol.~01, 2023, pp. 535--542.

\bibitem{Nurse_Scheduling_QA}
K.~Ikeda, Y.~Nakamura, and T.~Humble, ``Application of quantum annealing to nurse scheduling problem,'' \emph{Scientific Reports}, vol.~9, 09 2019.

\bibitem{youtube_encoding_settings}
Google, ``Recommended upload encoding settings,'' \url{https://support.google.com/youtube/answer/1722171?hl=en#zippy=%2Cbitrate}.

\bibitem{openqaoasdkqaoa}
\BIBentryALTinterwordspacing
V.~Sharma, N.~S.~B. Saharan, S.-H. Chiew, E.~I.~R. Chiacchio, L.~Disilvestro, T.~F. Demarie, and E.~Munro, ``Openqaoa -- an sdk for qaoa,'' 2022. [Online]. Available: \url{https://arxiv.org/abs/2210.08695}
\BIBentrySTDinterwordspacing

\bibitem{dwave_documentation}
\BIBentryALTinterwordspacing
{D-Wave Systems}, \emph{D-Wave Documentation}, 2023, accessed: 2024-09-21. [Online]. Available: \url{https://docs.dwavesys.com/docs/latest/index.html}
\BIBentrySTDinterwordspacing

\bibitem{UGCDataset}
Y.~Wang, S.~Inguva, and B.~Adsumilli, ``Youtube ugc dataset for video compression research,'' in \emph{2019 IEEE 21st International Workshop on Multimedia Signal Processing (MMSP)}, 2019, pp. 1--5.

\bibitem{NetflixDataset}
\BIBentryALTinterwordspacing
Netflix, ``Netflix public dataset for vmaf,'' 2016. [Online]. Available: \url{https://github.com/Netflix/vmaf/blob/master/resource/doc/datasets.md}
\BIBentrySTDinterwordspacing

\bibitem{AVADataset}
C.~Gu, C.~Sun, D.~A. Ross, C.~Vondrick, C.~Pantofaru, Y.~Li, S.~Vijayanarasimhan, G.~Toderici, S.~Ricco, R.~Sukthankar, C.~Schmid, and J.~Malik, ``Ava: A video dataset of spatio-temporally localized atomic visual actions,'' in \emph{2018 IEEE/CVF Conference on Computer Vision and Pattern Recognition}, 2018, pp. 6047--6056.

\end{thebibliography}

\end{document}